\documentclass[%
 reprint,
superscriptaddress,
 amsmath,amssymb,
 aps,
 prd,
floatfix,
showpacs
]{revtex4-2}

\pdfoutput=1

\usepackage{graphicx}
\usepackage{dcolumn}
\usepackage{bm}
\usepackage{lineno}
\newcommand{\dzeta}{\Delta\langle\zeta\rangle}
\newcommand{\kt}{\langle k_T \rangle}
\newcommand{\ktu}{\langle k_T^{u} \rangle}
\newcommand{\ktd}{\langle k_T^{d} \rangle}

\newcommand{\ktgsea}{\langle k_T^{g+sea}\rangle}
\newcommand{\etatot}{\eta^{total}}

\begin{document}

\title{Measurement of transverse single-spin asymmetries for dijet production in polarized proton-proton collisions at $\sqrt{s} = 200$ $\mathrm{GeV}$}

\affiliation{Academia Sinica, Nankang, 115}
\affiliation{Abilene Christian University, Abilene, Texas   79699}
\affiliation{Alikhanov Institute for Theoretical and Experimental Physics NRC "Kurchatov Institute", Moscow 117218}
\affiliation{Argonne National Laboratory, Argonne, Illinois 60439}
\affiliation{American University in Cairo, New Cairo 11835, Egypt}
\affiliation{Ball State University, Muncie, Indiana, 47306}
\affiliation{Brookhaven National Laboratory, Upton, New York 11973}
\affiliation{University of Calabria \& INFN-Cosenza, Rende 87036, Italy}
\affiliation{University of California, Berkeley, California 94720}
\affiliation{University of California, Davis, California 95616}
\affiliation{University of California, Los Angeles, California 90095}
\affiliation{University of California, Riverside, California 92521}
\affiliation{Central China Normal University, Wuhan, Hubei 430079 }
\affiliation{University of Illinois at Chicago, Chicago, Illinois 60607}
\affiliation{Chongqing University, Chongqing, 401331}
\affiliation{Creighton University, Omaha, Nebraska 68178}
\affiliation{Czech Technical University in Prague, FNSPE, Prague 115 19, Czech Republic}
\affiliation{National Institute of Technology Durgapur, Durgapur - 713209, India}
\affiliation{ELTE E\"otv\"os Lor\'and University, Budapest, Hungary H-1117}
\affiliation{Frankfurt Institute for Advanced Studies FIAS, Frankfurt 60438, Germany}
\affiliation{Fudan University, Shanghai, 200433 }
\affiliation{Guangxi Normal University, Guilin, 541004}
\affiliation{University of Heidelberg, Heidelberg 69120, Germany }
\affiliation{University of Houston, Houston, Texas 77204}
\affiliation{Huzhou University, Huzhou, Zhejiang  313000}
\affiliation{Indian Institute of Science Education and Research (IISER), Berhampur 760010 , India}
\affiliation{Indian Institute of Science Education and Research (IISER) Tirupati, Tirupati 517507, India}
\affiliation{Indian Institute Technology, Patna, Bihar 801106, India}
\affiliation{Indiana University, Bloomington, Indiana 47408}
\affiliation{Institute of Modern Physics, Chinese Academy of Sciences, Lanzhou, Gansu 730000 }
\affiliation{University of Jammu, Jammu 180001, India}
\affiliation{Joint Institute for Nuclear Research, Dubna 141 980}
\affiliation{Kent State University, Kent, Ohio 44242}
\affiliation{University of Kentucky, Lexington, Kentucky 40506-0055}
\affiliation{Lanzhou University, Lanzhou, 730000}
\affiliation{Lawrence Berkeley National Laboratory, Berkeley, California 94720}
\affiliation{Lehigh University, Bethlehem, Pennsylvania 18015}
\affiliation{Lovely Professional University, Jalandhar - Delhi G.T. Road, Pagwara, Panjab, 144411, India}
\affiliation{Max-Planck-Institut f\"ur Physik, Munich 80805, Germany}
\affiliation{Michigan State University, East Lansing, Michigan 48824}
\affiliation{National Research Nuclear University MEPhI, Moscow 115409}
\affiliation{National Institute of Science Education and Research, HBNI, Jatni 752050, India}
\affiliation{National Cheng Kung University, Tainan 70101 }
\affiliation{The Ohio State University, Columbus, Ohio 43210}
\affiliation{Panjab University, Chandigarh 160014, India}
\affiliation{NRC "Kurchatov Institute", Institute of High Energy Physics, Protvino 142281}
\affiliation{Purdue University, West Lafayette, Indiana 47907}
\affiliation{Rice University, Houston, Texas 77251}
\affiliation{Rutgers University, Piscataway, New Jersey 08854}
\affiliation{University of Science and Technology of China, Hefei, Anhui 230026}
\affiliation{South China Normal University, Guangzhou, Guangdong 510631}
\affiliation{Sejong University, Seoul, 05006, Korea, Republic Of}
\affiliation{Shandong University, Qingdao, Shandong 266237}
\affiliation{Shanghai Institute of Applied Physics, Chinese Academy of Sciences, Shanghai 201800}
\affiliation{Southern Connecticut State University, New Haven, Connecticut 06515}
\affiliation{State University of New York, Stony Brook, New York 11794}
\affiliation{Instituto de Alta Investigaci\'on, Universidad de Tarapac\'a, Arica 1000000, Chile}
\affiliation{Temple University, Philadelphia, Pennsylvania 19122}
\affiliation{Texas A\&M University, College Station, Texas 77843}
\affiliation{Texas Southern University, Houston, Texas, 77004}
\affiliation{University of Texas, Austin, Texas 78712}
\affiliation{Tsinghua University, Beijing 100084}
\affiliation{University of Tsukuba, Tsukuba, Ibaraki 305-8571, Japan}
\affiliation{University of Chinese Academy of Sciences, Beijing, 101408}
\affiliation{Valparaiso University, Valparaiso, Indiana 46383}
\affiliation{Variable Energy Cyclotron Centre, Kolkata 700064, India}
\affiliation{Warsaw University of Technology, Warsaw 00-661, Poland}
\affiliation{Wayne State University, Detroit, Michigan 48201}
\affiliation{Wuhan University of Science and Technology, Wuhan, Hubei 430065}
\affiliation{Yale University, New Haven, Connecticut 06520}

\author{B.~E.~Aboona}\affiliation{Texas A\&M University, College Station, Texas 77843}
\author{J.~Adam}\affiliation{Czech Technical University in Prague, FNSPE, Prague 115 19, Czech Republic}
\author{G.~Agakishiev}\affiliation{Joint Institute for Nuclear Research, Dubna 141 980}
\author{I.~Aggarwal}\affiliation{Panjab University, Chandigarh 160014, India}
\author{M.~M.~Aggarwal}\affiliation{Panjab University, Chandigarh 160014, India}
\author{Z.~Ahammed}\affiliation{Variable Energy Cyclotron Centre, Kolkata 700064, India}
\author{A.~Aitbayev}\affiliation{Joint Institute for Nuclear Research, Dubna 141 980}
\author{I.~Alekseev}\affiliation{Alikhanov Institute for Theoretical and Experimental Physics NRC "Kurchatov Institute", Moscow 117218}\affiliation{National Research Nuclear University MEPhI, Moscow 115409}
\author{E.~Alpatov}\affiliation{National Research Nuclear University MEPhI, Moscow 115409}
\author{A.~K.~Alshammri}\affiliation{Kent State University, Kent, Ohio 44242}
% ...

\collaboration{STAR Collaboration}\noaffiliation

\begin{abstract}
We report a new measurement of transverse single-spin asymmetries for dijet production in collisions of polarized protons at $\sqrt{s}$ = 200 $\mathrm{GeV}$.
Correlations between the proton spin 
and the transverse momenta of its partons, 
each perpendicular to the proton momentum direction, are probed at
high $Q^2$ $\approx $160 $\mathrm{GeV}^2$.
Evidence for 
nonzero Sivers effects is measured
for the first time in dijets from proton-proton
collisions, but only when the jets are sorted by their net charge, which enhances the
otherwise canceling opposite-sign
$u$- or $d$-quark contributions to separate data samples.
The resulting asymmetries are compared to recent theoretical calculations.
Separately, the associated Sivers observable $\kt$, the average parton transverse momentum, is extracted using a simple kinematics approach which further enables a determination of the individual partonic contributions to the observed asymmetries.
\end{abstract}


\maketitle

Our understanding of the three-dimensional structure of the nucleon has progressed significantly in the past decades~\cite{nucleon_structure,ANSELMINO2020103806} and will be further advanced at the future Electron-Ion Collider~\cite{EIC}.
In momentum space, nucleon structure is typically expressed via transverse-momentum-dependent parton distribution functions (TMD PDFs) with explicit dependence on the 
partonic transverse momentum ($\vec k_T$).
One TMD PDF of particular interest is the spin-dependent Sivers function~\cite{sivers}  $f^{\perp}_{1T}$ which characterizes 
a scalar triple-vector correlation for an unpolarized parton and its transversely polarized parent proton: $(\vec{k}_T \times \vec{S})\cdot \vec{P}$, where 
$\vec{S}$ and $\vec{P}$ are the 
proton spin and proton momentum, respectively.
In the hard scattering of transversely polarized protons, this correlation leads to a left-right asymmetry in the azimuthal distribution of produced particles. The Sivers effect was originally introduced~\cite{sivers} to explain the large transverse single-spin asymmetries (TSSA) observed in inclusive pion production, whose persistence to high transverse momenta $p_T$ appeared contrary to quantum chromodynamics (QCD) expectations~\cite{PhysRevLett.41.1689,PhysRevD.103.092009}. 
Currently, in addition to the TMD PDF and fragmentation function framework that
has expanded to describe the growing number of TMD phenomena, a 
collinear formalism involving twist-3 distributions 
(quark-gluon-quark correlations) continues its 
development and is
more applicable for describing TSSA effects in inclusive single-hadron production.
 The Sivers function and its twist-3 analog, the Efremov-Teryaev-Qiu-Sterman
distribution~\cite{PhysRevD.59.014004,Efremov:1981sh}, are related quantitatively~\cite{BOER2003201}, providing 
constraints and insight from different kinematic regimes.

Experimental evidence for the Sivers effect was first obtained~\cite{PhysRevLett.94.012002}
in semi-inclusive deep inelastic scattering (SIDIS)~\cite{hermesSIDIS, jlabSIDIS,compassSIDIS,PhysRevC.89.042201,ADOLPH2015250,Airapetian2020,PhysRevLett.133.101903}. Fits to these data show opposite signs and similar scale for the $u$- and $d$-quark Sivers functions~\cite{PhysRevD.72.094007,PhysRevD.72.054028,PhysRevD.73.014021, Anselmino_2008,PhysRevLett.107.212001,PhysRevD.88.114012,PhysRevD.89.074013,Anselmino_2017,
BOGLIONE2021136135,BACCHETTA2022136961},
with sea quarks compatible with zero.
Valence quarks nearly saturate the Burkardt sum rule~\cite{PhysRevD.69.091501}, leaving little room for gluon Sivers contributions.
Building on an unpolarized TMD foundation~\cite{Bacchetta2022},
there is also considerable recent interest in combining TMD data from SIDIS, $e^+e^-$ annihilation and ${pp}$ scattering to arrive at a unified 
picture including effects of the Sivers 
function~\cite{PhysRevD.102.054002,PhysRevLett.126.112002,BOGLIONE2021136135,
Echevarria_2021,BACCHETTA2022136961}.
While there is as yet no formal connection between the Sivers function and orbital angular momentum (OAM),  the latter is a prerequisite~\cite{BRODSKY200299}
for the Sivers effect. 
These and other studies~\cite{PhysRevLett.107.212001} point to an emerging nucleon
3-D structure and further understanding of possible
contributions of OAM to the nucleon spin.   

A distinctive feature of the Sivers function is its nonuniversality.
QCD gauge invariance requires the Sivers function to be process dependent, 
a manifestation of the underlying color dynamics, 
resulting in opposite signs for the Sivers asymmetries
in SIDIS and the Drell-Yan process~\cite{signchange}.
Investigations are ongoing
to confirm this predicted sign change
using $W^{\pm}/Z^{0}$ boson production
~\cite{doi:10.1146/annurev-nucl-102014-021948}, with only 
qualitative support so far observed~\cite{compassDY,2016sign,PhysRevLett.126.112002}. 
Recent COMPASS Drell-Yan results obtained via virtual photons~\cite{PhysRevLett.133.071902} indicate a consistency with the QCD sign change prediction.

Dijet production TSSAs in transversely polarized proton-proton collisions
are directly sensitive to the Sivers functions of the participating partons 
and can be experimentally studied via
the dijet opening-angle kinematic tilt (from back-to-back) that reverses
under spin-flip of the proton beam~\cite{zetadef}.
Use of the dijet channel avoids spin-correlated fragmentation contributions
along with suppression of some potential background effects, and at STAR probes
a significantly higher $Q^2$ scale ($\gtrsim$ 160 $\mathrm{GeV}^2$) than
previous Sivers effect studies. While the resulting asymmetries involve
gauge link contributions associated with color flow to both the initial
(as in Drell-Yan) and final (as in SIDIS) states~\cite{doi:10.1063/1.2750844},
when compared with theory, these measurements can help probe and
elucidate novel aspects of the QCD dynamics. 
An additional significant feature is that dijet asymmetry
measurements offer a means to directly explore the underlying spin-dependent
partonic $\vec k_T$ that characterizes the Sivers effect, offering the potential
for new insights regarding this important transverse momentum component.

An early analysis~\cite{2006dijet} with limited statistics from STAR at the
Relativistic Heavy Ion Collider (RHIC) found Sivers asymmetries consistent with
zero in dijet production, presumably due to the cancellation between opposite-sign
$u$ and $d$ quark contributions, and as suggested by then-current 
calculations~\cite{PhysRevD.75.074019}. 
However, deciphering results from back-to-back dijets in a theoretical framework faces nontrivial
issues from TMD factorization breaking~\cite{PhysRevD.72.034030,PhysRevD.75.114014,
PhysRevD.81.094006}, largely due to their two-scale (jet vs. dijet momentum balance)
nature. Among the challenges are 
resummation of large
logarithmic terms as well as the spin-dependent treatment of global soft functions.
New Sivers dijet measurements may 
enable additional insights and there are recent 
theoretical approaches spurred by such 
interest~\cite{PhysRevD.102.114012,2021JHEP...02..066K}.
We revisit the Sivers dijet measurement at STAR with a novel jet-charge tagging method to help separate the $u$ and $d$ contributions, together with significantly
improved statistics. Increased precision arises from inclusion of 
charged particle tracking in jet reconstruction.

In this analysis, we use 200 $\mathrm{GeV}$ transversely polarized 
$pp$ data collected in 2012 and 2015 at STAR, 
corresponding to integrated luminosities of 22 and 52 pb$^{-1}$, respectively. The involved subsystems of the STAR detector~\cite{STAR_Detector} 
are the Time Projection Chamber (TPC)~\cite{STAR_TPC}, 
providing charged particle tracking for pseudorapidity 
$|\eta^{detector}| \leq$ 1.3, and the Electromagnetic Calorimeter (EMC), 
measuring the energy of electrons and photons while providing jet triggering
in the barrel \break
-1 $<\eta^{detector}<$ 1 (BEMC~\cite{STAR_EMC}) and endcap 1.1 $<\eta^{detector}<$ 2 (EEMC~\cite{ALLGOWER2003740}) regions with full azimuthal ($\phi$) coverage.
The polarizations for the 
circulating beams are measured using Coulomb-nuclear interference
proton-carbon polarimeters~\cite{JINNOUCHI_2005}, calibrated with a polarized hydrogen gas-jet target~\cite{OKADA2006450}. 
The average beam polarization magnitudes are 56\% (2012) and 57\% (2015), 
both with a relative scale uncertainty of 
3.2\%~\cite{STAR_Polarimetry}. The analysis treats only one beam as polarized 
at a time.

The data are selected using an EMC jet-patch trigger with two levels of
transverse energy ($E_T$) threshold in 
a $\Delta\eta \times \Delta\phi = 1\times$1 ($\mathrm{radians}$) region: 5.4 $\mathrm{GeV}$ (JP1) and 7.3 $\mathrm{GeV}$ (JP2). 
Jets are reconstructed using the anti-$k_{T}$~\cite{Matteo_Cacciari_2008} algorithm 
with R = $\sqrt{\Delta\phi^2+\Delta\eta^2}$ = 0.6,
employed with standard STAR selection criteria on the TPC tracks, EMC towers and 
proto-jet quantities~\cite{2009ALL}. To ensure the quality of the dijets, we select events with exactly two jets, 
one having $p_T$ $>$ 6 $\mathrm{GeV/c}$ and the other $p_T$ $>$ 4 $\mathrm{GeV/c}$, 
with a relative opening angle $\Delta\phi$ $>$ 120$^{\circ}$. 
Both jets are required to originate from a single vertex with $|\mathrm{Z}_{vertex}|$ $<$ 90 cm, 
and fall within -0.8 $<$ $\eta$ $<$ 1.8 and -0.7 $<$ $\eta^{detector}$ $<$ 1.7.
In order to avoid false triggering effects, a trigger simulator is applied, which
requires the matching of offline reconstructed jets with triggered jet patches. The resulting number of events is $\sim$33 times
that available in an earlier study~\cite{2006dijet}.

The observable in this analysis relies on precise knowledge of the jet
directions (as compared to the magnitude of their momenta),
the \textit{signed dijet opening angle} ($\zeta$)~\cite{zetadef}. It is defined as 
$\zeta$ = $|\Delta\phi|$ if cos($\phi_{b}$) $<$ 0 and 
$\zeta$ = 360$^{\circ}
- |\Delta\phi|$ if cos($\phi_{b}$) $>$ 0, 
where $\phi_{b}$ is the azimuthal angle of the bisector ray, which reverses direction when the beam polarization direction is flipped.
The sensitivity of $\zeta$ to transverse spin effects 
is maximized when the two jets are parallel to the beam spin 
and, modulated by $|$cos($\phi_{b}$)$|$, is zero when the two jets are perpendicular. 

Our method to extract an asymmetry for the spin-dependent dijet response differs, 
but is closely related to the 
single-spin analyzing power $A_N$.
We choose the asymmetry calculated as the
difference of $\zeta$ centroids
$\langle\zeta\rangle$ between the spin-up and spin-down states:
\begin{equation}
    \dzeta = \frac{\langle\zeta\rangle^{+} - \langle\zeta\rangle^{-}}{P},
    \label{eq:asymm}
\end{equation}
where $\langle\zeta\rangle^{+/-}$ is the centroid of the Gaussian-like~\cite{2006dijet} $\zeta$ distribution 
in the spin-up/spin-down state, and $P$ is the beam polarization. 
The asymmetry in Eq.~(\ref{eq:asymm}) is motivated by the fact that for our jet $p_T$ range, $\zeta$ is directly sensitive to the Sivers function $\kt$, driver of the 
spin-dependent shift in the dijet opening angle. 
This construction also sidesteps
several systematic uncertainties, such as relative luminosity, asymmetric detector azimuthal acceptance and background contribution effects.
$\langle\zeta\rangle$ is extracted by fitting the 
$\zeta$ distribution $N(\zeta)$
over a selected range with a three-Gaussian function
capturing its salient features: 
\begin{equation}
    N(\zeta) =
    p_0 \cdot (e^{-\frac{{(\zeta - p_1)}^2}{2 {p_2}^2}} 
    + p_3 \cdot e^{-\frac{{(\zeta - p_1)}^2}{2 {p_4}^2}}
    + p_5 \cdot e^{-\frac{{(\zeta - p_1)}^2}{2 {p_6}^2}}),
    \label{eq:3gaus}
\end{equation}
where all components share the same 
peak position $p_1$, taken as $\langle\zeta\rangle$. 
Values of the centroid differences $\dzeta$ 
subsequently extracted are largely insensitive to variation of this empirically driven shape~\cite{supplement}.
Two fitting steps are performed: 
1) spin-up and spin-down distributions, scaled to the same integral, are combined and fit to determine  
individual Gaussian parameters;  
2) spin-up and spin-down $\zeta$ distributions are separately fit using Eq.~(\ref{eq:3gaus}), with only $p_1$ allowed to vary,
making the final fit results more sensitive to 
the distribution centroid and improving accuracy
and stability. The broad $\zeta$ distribution, driven mainly by parton-level 
multi-$\mathrm{GeV}$ intrinsic 
${k_T}$ and initial-state radiation effects~\cite{perugia}, nearly fills 
the analyzed back-to-back dijet $\zeta$ range of 180$\pm$60$^{\circ}$; our fit range of 180$\pm$50$^{\circ}$ allows some room for systematic study.
The resulting $\dzeta$ vs. $\phi_b$ values are 
fit with a cosine function, whose amplitude quantifies 
the measured asymmetry.

\begin{figure}[htb!]
    \vspace{-5pt}
    \centering
    \includegraphics[width=7cm]{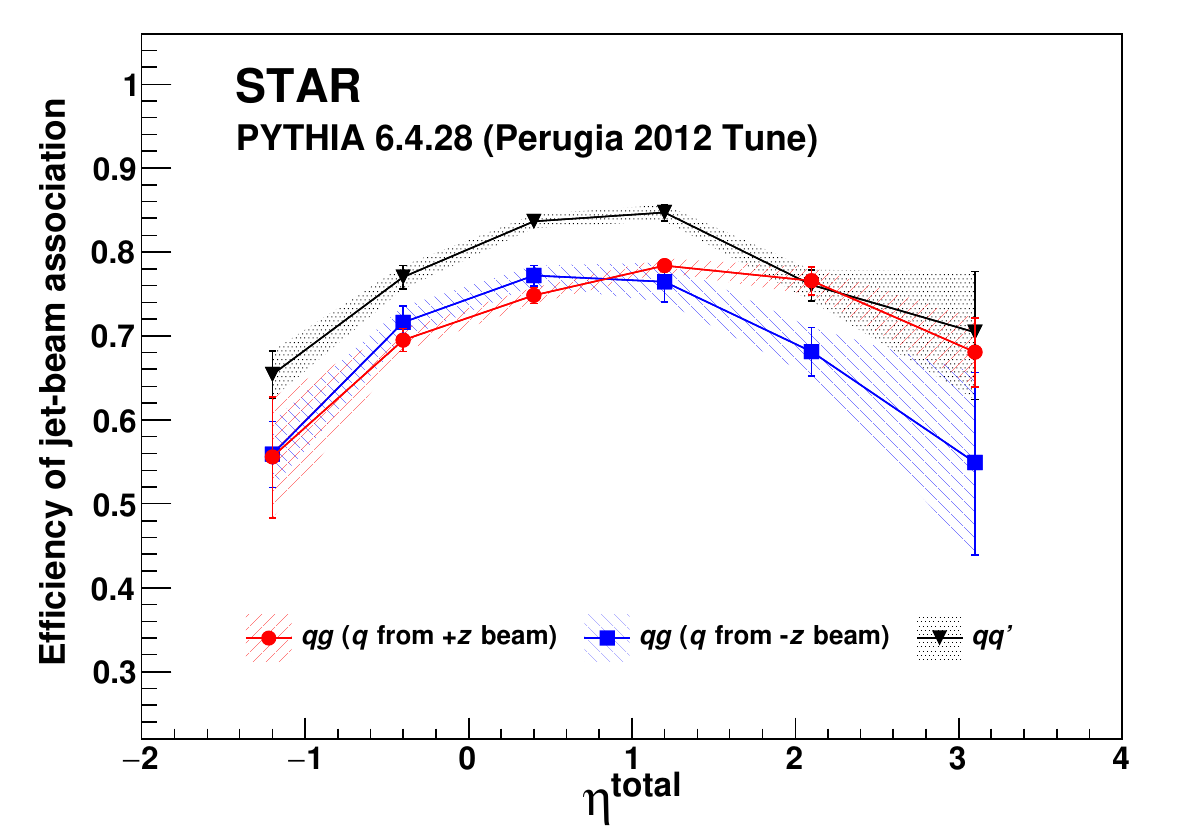}
    \caption{Efficiency for associating beam and jet correctly in the embedding sample vs.  $\eta^{total}$, the summed dijet $\eta$. 
    }
    \vspace{-5pt}
    \label{jet_assoc}
\end{figure}

To avoid possible $u$-$d$ cancellation, 
we divide the data into multiple kinematic regions in the analysis, each with a variant parton composition, 
with
the $\dzeta$ asymmetry extracted separately in each of four ``jet charge" bins within the subsets of JP triggers and 2012, 2015 data.
The initial selection is based on ``tagging" of a jet, which 
must first be associated with the polarized beam. 
The more forward of the jets (largest detector $|\eta|$) is assumed to be likely coming from the scattered parton of the beam pointing into the same hemisphere. 
For instance, in a dijet event with $\eta_1 > \eta_2$, 
$jet_1$ is associated with the +$z$ beam and $jet_2$ with the $-z$ beam.
Performance of the beam-jet association is studied with simulation based on Pythia 6.4.28~\cite{pythia} (Perugia 2012 tune~\cite{perugia,Perugia_2012}) 
and GEANT 3~\cite{geant}, and embedded into randomly selected beam bunch crossings from data to mimic real beam background, pileup and detector inefficiencies. 
Simulation reveals that the resulting association efficiency for $qg$ and $qq'$ subprocesses averages about 70\%-75\% (Fig.~\ref{jet_assoc}). 
This ensures good performance of jet tagging for the $u$ and $d$ 
quarks in the next step. (Note: for identical partons, $gg$ and $qq$, the association is ambiguous.)

\begin{figure}[hb!]
    \vspace{-10pt}
    \centering
    \includegraphics[width=7cm]{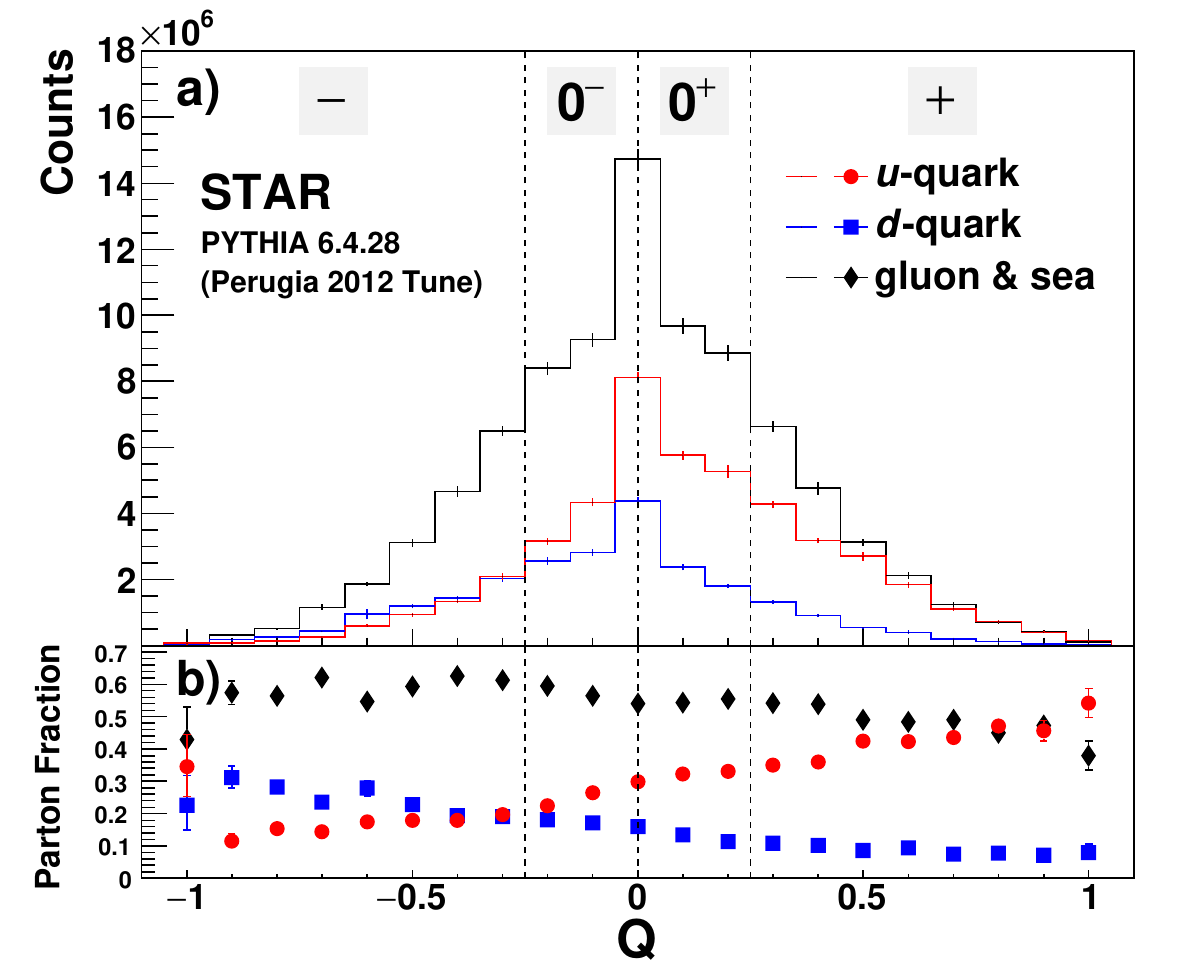}
    \vspace{-1pt}
    \caption{The (a) distribution of $Q$ and (b) respective parton fraction from the embedded simulation. 
    The tagging divides the data sample into 4 bins (separated by the dashed lines).} 
    \vspace{-5pt}
    \label{Q_charge}
\end{figure}

\begin{figure}[htb!]
    \vspace{-0pt}
    \centering
    \includegraphics[width=8cm]{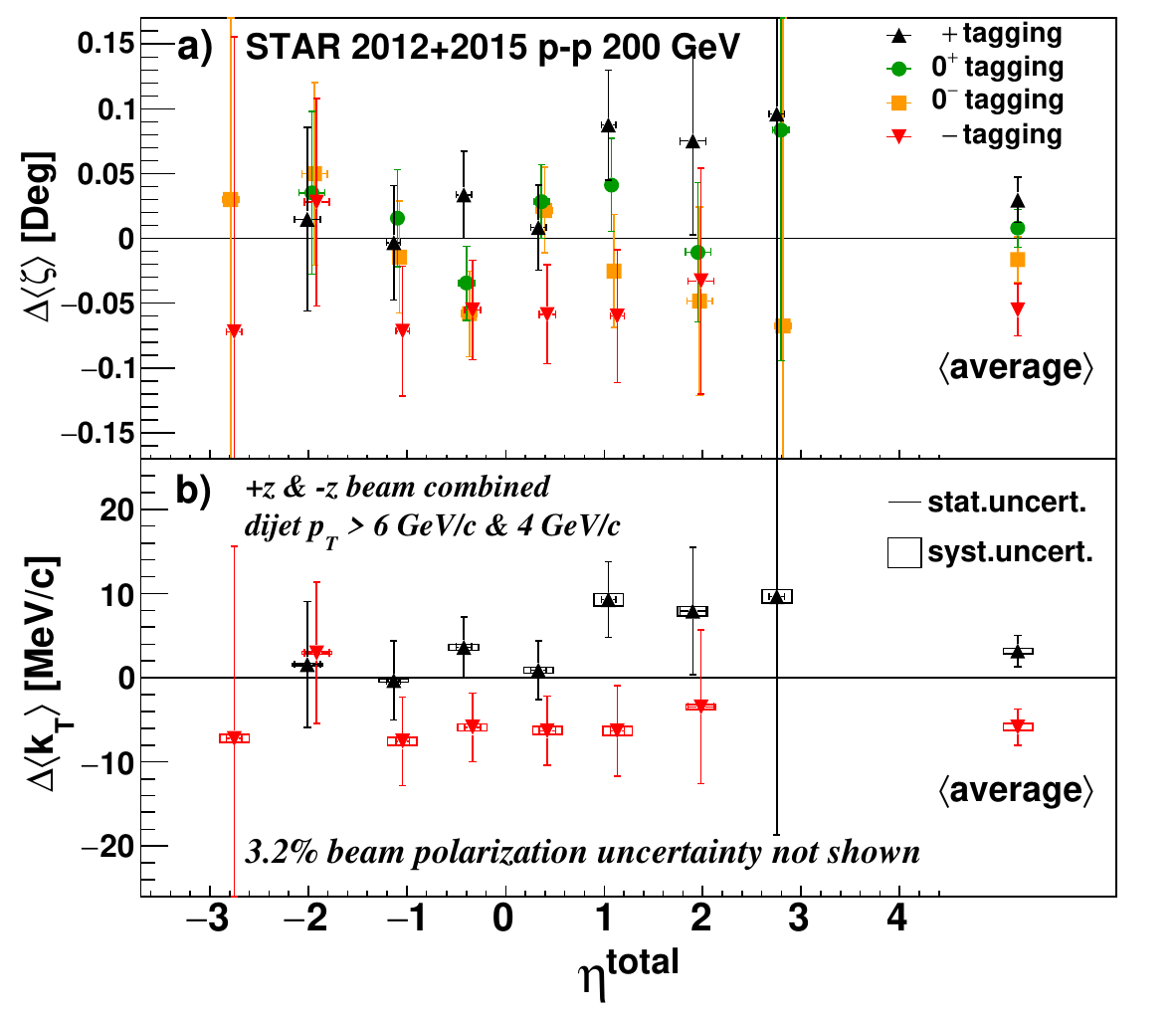}
    \vspace{-1pt}
    \caption{The (a) $\dzeta$ values and (b) converted $\Delta\kt$ plotted as a function of $\etatot$.  
    Rightmost points represent the
    average over the $\etatot$ bins. The $0^+$ and $0^-$ points are suppressed in the lower panel  
    to better view
    the $\Delta\kt$ signal and systematic errors (dominated by fitting range). 
    Plotted points are offset in $\etatot$ and outsize values omitted for clarity.}
    \vspace{-10pt}
    \label{dZeta_kT}
\end{figure}

During hadronization, the $u$ quarks and $d$ quarks produce relatively 
more positively charged and negatively charged particles, respectively.
This feature can be quantified by jet charge ($Q$)~\cite{jetcharge} to help in tagging jets:
\begin{equation}
    Q = \sum_{|p^{track}| > 0.8\; GeV/c} \frac{|p^{track}|}{|p^{jet}|} \cdot q^{track}  \;, 
    \label{eq:qsum}
\end{equation}
where $q^{track}$ is the charge sign of each track.
To reduce the influence from underlying events, 
only tracks with $|p|$ $>$ 0.8 $\mathrm{GeV/c}$ are selected in the calculation. The distributions of $Q$ for different scattered partons (gluon and antiquark sea combined) are plotted 
using the embedding sample in Fig.~\ref{Q_charge}, 
for which the effect
of beam-jet association has also been folded in. 
Based on these plots, each data sample is divided into four ``jet-charge" bins:
\begin{itemize}
    \setlength\itemsep{0em}
    \item + tagging: $Q$ $\geq$ 0.25, enhancing the fraction of $u$
    \item $0^{+}$ tagging: 0 $\leq$ $Q$ $<$ 0.25, less enhancement of $u$
    \item $0^{-}$ tagging: -0.25 $<$ $Q$ $<$ 0, less enhancement of $d$
    \item $-$ tagging: $Q$ $\leq$ -0.25, enhancing the fraction of $d$
\end{itemize}
The four binned regions are expected to show different $\dzeta$ asymmetries 
assuming opposite signs of the Sivers function for the $u$ and $d$ quarks.
Since the parton fraction is dependent on Bjorken $x$, we further divide the 
full data set into $\etatot$ bins ($\etatot$ = $\eta_1+\eta_2$ $\propto$ 
$\log(x_1/x_2)$)~\cite{2006dijet}. We also combine the separate $+z$ 
and $-z$ polarized beam results, by transforming 
(rotation around the $y$-axis) the $-z$
beam $\dzeta$ asymmetries into the $+z$ beam direction.
The resulting $\dzeta$ asymmetries, now over an extended
$\etatot$ range, for the 
four charge-tagged bins are shown in Fig.~\ref{dZeta_kT} (a).
We observe an $3.1\sigma$
separation between the averaged asymmetries 
for the +~tagging and the $-$~tagging.
A correlation between $\dzeta$ and $Q$ is also manifest, as the asymmetry shifts from negative to positive with increasing $Q$ (less $d$ and more $u$).
This is strong evidence that the Sivers $\kt$ effect for $u$ and $d$ are opposite in sign, as indicated in SIDIS measurements~\cite{hermesSIDIS,compassSIDIS,jlabSIDIS}.

Our measured $\dzeta$ asymmetries are validated through several crosschecks.
A null test made by calculating the asymmetry in the direction orthogonal to the 
expected Sivers $\vec{k_T}$ finds all the charge-tagged results consistent with zero,
ruling out the possibility of major spin-dependent systematic effects.
In the separated +$z$ beam and $-z$ beam measurements, 
we see overall consistency in sign and magnitude for the asymmetries within 
the same $\etatot$ bins and the same charge-tagged bins.
Similar agreement is also observed for the results using only 2012 or 2015 data. 

\begin{figure}[hbt!]
    \vspace{-5pt}
    \centering
    \includegraphics[width=8.6cm]{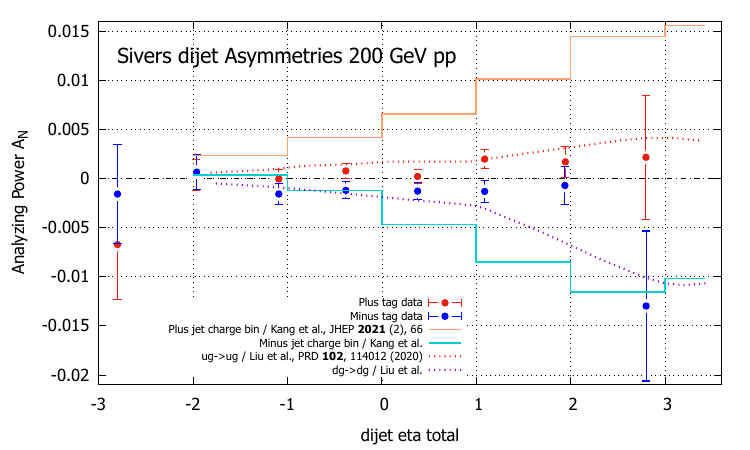}
    \vspace{-10pt}
    \caption{Comparisons of $A_N$ for 200 $\mathrm{GeV}$ $pp$ Sivers dijet asymmetries with theoretical predictions from Kang \textit{et al.}~\cite{2021JHEP...02..066K} and Liu \textit{et al.}~\cite{PhysRevD.102.114012}. Measured $\dzeta$ asymmetries have been converted to $A_N$ via a numerical factor as described in the text (a 10\% per point RMS conversion error is not included).}   
    \vspace{-15pt}
    \label{theory_compare}
\end{figure}

Recent theoretical calculations~\cite{2021JHEP...02..066K,PhysRevD.102.114012}
 of the Sivers asymmetries 
for 200 $\mathrm{GeV}$ $pp$ dijets are expressed in terms of the analyzing power observable $A_N$.
The connection between $\dzeta$ of Eq.~(\ref{eq:asymm}) and $A_N$ is made using
the dijet opening angle $\zeta$ distribution itself, with $A_N$ characterizing
the left-right event difference arising from the tilt of nominal back-to-back jets~\cite{zetadef},
while $\dzeta$ is a measure of the shift of the whole distribution.  
Since our $\dzeta$ centroid analysis involves Gaussian fitting of the 
$\zeta$ event distribution (total events), whose spin-dependent centroid shifts are small (hence the asymmetry driving events can be simply evaluated as the peak amplitude times $\dzeta$),
we can numerically~\cite{supplement} 
relate our $\dzeta$ results to those of a
$A_N$-focused analysis: e.g., a single Gaussian of width $\sigma$ gives $A_N = \frac{ampl\cdot\dzeta}{ampl\cdot \sigma\cdot\sqrt{2\pi}} = \frac{\dzeta}{\sigma\cdot\sqrt{2\pi}}.$
A comparison of such extracted $A_N$ values, for the tagged results of Fig.~\ref{dZeta_kT} (a) with theoretical calculations is shown in 
Fig.~\ref{theory_compare}. 

Overall, one sees that both calculations 
and the data are comparable in magnitude, and  
correlate with the sign of the 
asymmetries for the plus ($u$ quark)- vs. minus ($d$ quark)-tagged data, generally lending support to the scale of our experimental results.
The curves from Kang \textit{et al.}~\cite{2021JHEP...02..066K} are based on a 
QCD formalism explicitly allowing for resummation of logarithmic terms and
represent a full calculation, including a sum over all contributing
color line terms (e.g., both SIDIS- and DY-like contributions). While 
$Q$-charge and jet threshold selections are chosen to be similar to our experimental analysis, 
the predicted $\etatot$ dependence is not seen in our data.
The calculations from Liu \textit{et al.}~\cite{PhysRevD.102.114012} on the other
hand, include only the use of one loop soft gluons and TMD factorization at 
leading order. The curves display just the behavior of the dominant $ug \rightarrow ug$ and 
$dg \rightarrow dg$ channels, and while they don't include tagging efficiency, they do
generally describe both the magnitude and more gentle $\etatot$ behavior of the data 
reasonably well. 

The nominal agreement of the single channel Liu \textit{et al.} calculation with data, along with the similar sign and magnitude correlation of the Kang \textit{et al.} curves, may be some indication that the $pp$ dijet asymmetries are
dominated by final-state color flow (e.g., a sign-change negative initial-state 
contribution appears not to dominate). 

As segue and further connection of our $\dzeta$ results to
theoretical expressions, we note schematically that if 
$d\sigma^{\pm}(\zeta)$ denotes an element of the spin-dependent dijet 
opening angle distribution (e.g., cast the cross section expressions similar 
to Eqs.~(3) and (10) 
of ref.~\cite{2021JHEP...02..066K} in a form differential in $\zeta$), one recovers a familiar structure for $A_N$, and a similar expression with a $\zeta$ weighted
distribution integrated over $\zeta$ for $\dzeta$ : $\Delta\langle\zeta \rangle = (\int{\zeta d\sigma^+} -\int{\zeta d\sigma^-}) / \int{d\sigma}$. That is, the difference of the distribution centroids that we measure can also be directly related to theory.

Returning now to the $\dzeta$ results of Fig.~\ref{dZeta_kT} (a), we discuss the 
extraction of the associated average Sivers $\kt$. 
Formally, the difference of the individual $\zeta$ distribution-weighted averages of ${k_T}$ associated with the two spin orientations is given by
$\langle k_T\rangle^+ - \langle k_T\rangle^-  
= (\int{k_T d\sigma^+} -\int{k_T d\sigma^-}) / \int{d\sigma}$. This difference 
is effectively the average Sivers $\langle k_T\rangle^{Sivers}$ that drives
our measured $\dzeta$ asymmetry.
We calibrate it using the dijet opening angle shift resulting from a parton level, 
transverse ``kick'' by correlating detector level $\dzeta$ and parton level $\kt$
(notably independent of the intrinsic $k_T$ characterization as verified by a dedicated simulation).
To do the conversion, we first correct the jet $p_T$ back to its parton level
based on machine learning using the embedded simulation sample.
We adopt the same algorithm, variables and training configuration as in a previous analysis~\cite{2009ALL},  
but now targeted toward parton $p_T$ instead of particle jet $p_T$. 
The weights from the training are applied to the jets in the real data to determine the actual $p_T$ distribution. Next, the $\dzeta$-$\kt$ correlation is constructed using kinematics alone. We independently 
add two opposite constant $k_T$ vectors, 
($k_T$, 0, 0) and (-$k_T$, 0, 0),
to the corrected jet $p_T$ to mimic dijet tilts 
in nominal ``spin-up" and ``spin-down" states, respectively.
A $\dzeta$ asymmetry at parton level is extracted following the above analysis 
procedure; the results are used at detector level with a systematic described further below. By assigning 5 different $k_T$ values 
in the range 1-20 $\mathrm{MeV/c}$ to the
added vectors, an experimentally determined linear relation between 
$\dzeta$ and an effective $\kt$ is observed individually for each $\etatot$ bin, which can be well fit with a slope: $\dzeta = slope \cdot \kt$.
Due to $p_T$ differences in $\etatot$ bins, the slope ranges from 9.26$^{\circ}\cdot c/\mathrm{GeV}$ 
in the mid-rapidity region to 9.97$^{\circ}\cdot c/\mathrm{GeV}$ in the more forward region.

The $\dzeta$ results are converted to $\Delta\kt$ results, Fig.~\ref{dZeta_kT} (b),
by applying the reverse of the above calculated slope, 
$\Delta\kt$ = $\dzeta/slope$.
The $\etatot$-averaged $\Delta\kt$ is found to be 3.2 $\pm$ 1.9 $\mathrm{MeV/c}$ for the +tagging bin, and -5.8 $\pm$ 2.1 $\mathrm{MeV/c}$
for the $-$tagging bin. 
The untagged asymmetry, -0.4 $\pm$ 0.9 $\mathrm{MeV/c}$, obtained from the error-weighted mean of the four charge-tagged bins, is consistent with zero.
We observe a $\sim$2$\sigma$ level linear $\etatot$-dependency in the 
+tagging $\kt$ results.
This is likely due mainly to the $x$-dependence of the parton fractions.

\begin{figure}[htb!]
    \vspace{-0pt}
    \centering
    \includegraphics[width=8.4cm]{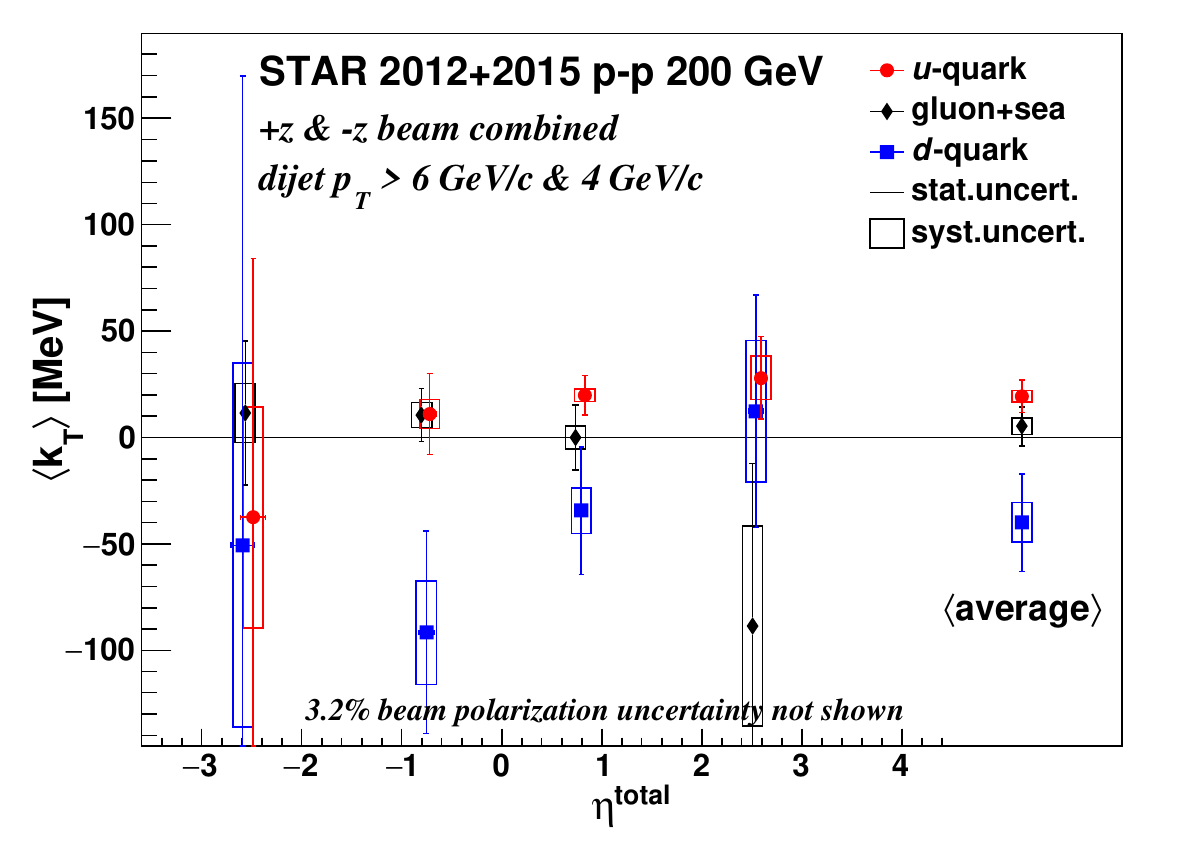}
    \vspace{-1pt}
    \caption{The $\kt$ for individual partons, inverted using parton fractions from 
    simulation and tagged $\kt$,  
    plotted as a function of $\etatot$, with rightmost points the $\etatot$ average. 
    Plotted points are offset in $\etatot$ for clarity,
    and systematic uncertainties in $\etatot$ are set nonzero to 
    improve visibility. 
    }
    \vspace{-5pt}
    \label{inverted_kT}
\end{figure}

The tagged $\Delta\kt$ results provide sufficient constraints to solve for the 
Sivers $\kt$ of individual partons if we know the parton fractions 
in each charge-tagged bin, which can be estimated from simulation. 
The inversion method used is based on fundamental concepts;  
we assume simple 2 $\rightarrow$ 2 scattering kinematics in the simulation 
sample which then guides the analysis to determine the magnitude and sign of
the Sivers $\kt$ for individual partons.
Combining the gluon and sea quark contributions, there are four constraints from charge tagging vs. three unknown variables: $\ktu$, $\ktd$ and $\ktgsea$.
To increase the stability of the inversion process, data from adjacent 
bins in $\etatot$ are combined, leading to the eight constraints:
\begin{equation}
    f_{i,j}^{u} \ktu + f_{i,j}^{d} \ktd + f_{i,j}^{g+sea} \ktgsea = \kt_{i,j}  \;,
    \label{eq:pseudoinv}
\end{equation}
where $f$ represents the parton fraction from simulation~\cite{supplement}, 
the right-hand side
$\kt$ is the tagged measurement in data, $i$ runs over all the charge tagging bins, 
and $j$ runs over the two adjacent $\etatot$ bins.
The over-constrained system is solved through Moore-Penrose inversion,
yielding values for the individual parton $\kt$, displayed in Fig.~\ref{inverted_kT} and discussed further below.

The systematic uncertainty of the parton $\kt$ 
has major contributions from two sources:
the fitting range of $\zeta$ and the more dominant error associated with the estimation of parton fractions. The fit range uncertainty is 
estimated by varying the range from
180$\pm$40$^{\circ}$ to 180$\pm$60$^{\circ}$ and  
extracting $\langle\zeta\rangle$ as a deviation from nominal; 
the resulting uncertainty is less than 15\% for  
+tagging/$-$tagging as indicated in Fig.~\ref{dZeta_kT} (b).
Parton fractions are estimated with leading-order PYTHIA simulations,
which come with their own set of contributing systematics, dominated by the choice of PDF and ISR/FSR (see Ref.~\cite{supplement}).
 These uncertainties vary with parton purity in the  
charge bins and as a function of 
$-3.6 < \etatot < 3.6$, ranging from 18 to 7-12\% for $u$ and $d$, and 3-21\% for $g+sea$. There is also a systematic associated with
the parton to detector level $\Delta\zeta$ mapping. 
A comparison of values for a wide range of simulated $\kt$, gives an 
uncertainty of $\sim$5.6\%, due largely 
to average momentum differences in the various bins.

The inverted results and average over all the $\etatot$ bins are shown in Fig.~\ref{inverted_kT}. The average $\ktu$ is estimated to be +19.3 $\pm$ 7.6 (stat.) $\pm$ 2.6 (syst.) $\mathrm{MeV/c}$, 
in which the positive sign means the $u$ quarks are correlated with the proton spin and proton momentum following the right-hand rule: 
$\vec{k_T^{u}} \cdot (\vec{S} \times \vec{P})$ $>$ 0.
In contrast, the average $\ktd$ is estimated to be 
-40.0 $\pm$ 22.9 $\pm$ 9.3 $\mathrm{MeV/c}$,
showing an opposite sign and a similar
magnitude compared to $\ktu$. While these results are likely somewhat
specific to their extraction in $pp$ dijets with possible contributions from
different color flow and an average over kinematic bins rather than global extaction, 
they are roughly comparable to the $u$-$d$ correlation in SIDIS measurements
at a much lower scale extracted as $96^{+60}_{-28}$ and 
$-113^{+45}_{-51}$ $\mathrm{MeV}$ for $u$ and $d$, respectively~\cite{Anselmino_2008}.   
We also find that the gluon and sea quarks are consistent with zero,
$\ktgsea$ = 5.3 $\pm$ 9.2 $\pm$ 3.8 $\mathrm{MeV/c}$.
Our measurement probes a range $0.03 < x < 0.6$ ($u$ and $d$ quarks), and  
$0.01 < x < 0.5$ (gluons); 
bin-by-bin parton $x$ and $\kt$ values are listed in Supplemental Material~\cite{supplement}.

In summary, transverse single-spin asymmetries for dijet production in $pp$ collisions are
measured in jet-charge bins using $\sqrt{s} =200~ \mathrm{GeV}$ data at STAR. 
This is the first time that evidence for nonzero Sivers signals in 
$pp$-induced dijet production is measured. Comparison of our results
with further theoretical developments may help to probe interesting aspects 
of the QCD dynamics. Through $\dzeta$-to-$\kt$ conversion and 
pseudoinversion, 
the Sivers $\kt$ for individual partons are unfolded in a kinematic
approach. The $u$- and $d$-quark $\kt$ are found to have opposite signs 
and similar magnitudes, while $\kt$ for gluon and sea quarks combined 
is consistent with zero within still large uncertainties.
Inclusion of these data in global analyses can
enhance a consistent extraction of Sivers observables and perhaps impact the  understanding of evolution effects, process dependence 
and other important issues relating to the Sivers TMD function.
   
We thank the RHIC Operations Group and SCDF at BNL, the NERSC Center at LBNL, and the Open Science Grid consortium for providing resources and support.  This work was supported in part by the Office of Nuclear Physics within the U.S. DOE Office of Science, the U.S. National Science Foundation, National Natural Science Foundation of China, Chinese Academy of Science, the Ministry of Science and Technology of China and the Chinese Ministry of Education, NSTC Taipei, the National Research Foundation of Korea, Czech Science Foundation and Ministry of Education, Youth and Sports of the Czech Republic, Hungarian National Research, Development and Innovation Office, New National Excellency Programme of the Hungarian Ministry of Human Capacities, Department of Atomic Energy and Department of Science and Technology of the Government of India, the National Science Centre and WUT ID-UB of Poland, German Bundesministerium f\"ur Bildung, Wissenschaft, Forschung and Technologie (BMBF), Helmholtz Association, Ministry of Education, Culture, Sports, Science, and Technology (MEXT), Japan Society for the Promotion of Science (JSPS), and Agencia Nacional de Investigacion y Desarrollo de Chile (ANID), Chile.

\bibliographystyle{apsrev4-2} 
\bibliography{main_text}

\end{document}


\title{Supplemental Material to: Measurement of transverse single-spin asymmetries for dijet production in polarized proton-proton collisions at $\sqrt{s} = 200$ GeV} 

\affiliation{Academia Sinica, Nankang, 115}
\affiliation{Abilene Christian University, Abilene, Texas   79699}
\affiliation{Alikhanov Institute for Theoretical and Experimental Physics NRC "Kurchatov Institute", Moscow 117218}
\affiliation{Argonne National Laboratory, Argonne, Illinois 60439}
\affiliation{American University in Cairo, New Cairo 11835, Egypt}
\affiliation{Ball State University, Muncie, Indiana, 47306}
\affiliation{Brookhaven National Laboratory, Upton, New York 11973}
\affiliation{University of Calabria \& INFN-Cosenza, Rende 87036, Italy}
\affiliation{University of California, Berkeley, California 94720}
\affiliation{University of California, Davis, California 95616}
\affiliation{University of California, Los Angeles, California 90095}
\affiliation{University of California, Riverside, California 92521}
\affiliation{Central China Normal University, Wuhan, Hubei 430079 }
\affiliation{University of Illinois at Chicago, Chicago, Illinois 60607}
\affiliation{Chongqing University, Chongqing, 401331}
\affiliation{Creighton University, Omaha, Nebraska 68178}
\affiliation{Czech Technical University in Prague, FNSPE, Prague 115 19, Czech Republic}
\affiliation{National Institute of Technology Durgapur, Durgapur - 713209, India}
\affiliation{ELTE E\"otv\"os Lor\'and University, Budapest, Hungary H-1117}
\affiliation{Frankfurt Institute for Advanced Studies FIAS, Frankfurt 60438, Germany}
\affiliation{Fudan University, Shanghai, 200433 }
\affiliation{Guangxi Normal University, Guilin, 541004}
\affiliation{University of Heidelberg, Heidelberg 69120, Germany }
\affiliation{University of Houston, Houston, Texas 77204}
\affiliation{Huzhou University, Huzhou, Zhejiang  313000}
\affiliation{Indian Institute of Science Education and Research (IISER), Berhampur 760010 , India}
\affiliation{Indian Institute of Science Education and Research (IISER) Tirupati, Tirupati 517507, India}
\affiliation{Indian Institute Technology, Patna, Bihar 801106, India}
\affiliation{Indiana University, Bloomington, Indiana 47408}
\affiliation{Institute of Modern Physics, Chinese Academy of Sciences, Lanzhou, Gansu 730000 }
\affiliation{University of Jammu, Jammu 180001, India}
\affiliation{Joint Institute for Nuclear Research, Dubna 141 980}
\affiliation{Kent State University, Kent, Ohio 44242}
\affiliation{University of Kentucky, Lexington, Kentucky 40506-0055}
\affiliation{Lanzhou University, Lanzhou, 730000}
\affiliation{Lawrence Berkeley National Laboratory, Berkeley, California 94720}
\affiliation{Lehigh University, Bethlehem, Pennsylvania 18015}
\affiliation{Lovely Professional University, Jalandhar - Delhi G.T. Road, Pagwara, Panjab, 144411, India}
\affiliation{Max-Planck-Institut f\"ur Physik, Munich 80805, Germany}
\affiliation{Michigan State University, East Lansing, Michigan 48824}
\affiliation{National Research Nuclear University MEPhI, Moscow 115409}
\affiliation{National Institute of Science Education and Research, HBNI, Jatni 752050, India}
\affiliation{National Cheng Kung University, Tainan 70101 }
\affiliation{The Ohio State University, Columbus, Ohio 43210}
\affiliation{Panjab University, Chandigarh 160014, India}
\affiliation{NRC "Kurchatov Institute", Institute of High Energy Physics, Protvino 142281}
\affiliation{Purdue University, West Lafayette, Indiana 47907}
\affiliation{Rice University, Houston, Texas 77251}
\affiliation{Rutgers University, Piscataway, New Jersey 08854}
\affiliation{University of Science and Technology of China, Hefei, Anhui 230026}
\affiliation{South China Normal University, Guangzhou, Guangdong 510631}
\affiliation{Sejong University, Seoul, 05006, Korea, Republic Of}
\affiliation{Shandong University, Qingdao, Shandong 266237}
\affiliation{Shanghai Institute of Applied Physics, Chinese Academy of Sciences, Shanghai 201800}
\affiliation{Southern Connecticut State University, New Haven, Connecticut 06515}
\affiliation{State University of New York, Stony Brook, New York 11794}
\affiliation{Instituto de Alta Investigaci\'on, Universidad de Tarapac\'a, Arica 1000000, Chile}
\affiliation{Temple University, Philadelphia, Pennsylvania 19122}
\affiliation{Texas A\&M University, College Station, Texas 77843}
\affiliation{Texas Southern University, Houston, Texas, 77004}
\affiliation{University of Texas, Austin, Texas 78712}
\affiliation{Tsinghua University, Beijing 100084}
\affiliation{University of Tsukuba, Tsukuba, Ibaraki 305-8571, Japan}
\affiliation{University of Chinese Academy of Sciences, Beijing, 101408}
\affiliation{Valparaiso University, Valparaiso, Indiana 46383}
\affiliation{Variable Energy Cyclotron Centre, Kolkata 700064, India}
\affiliation{Warsaw University of Technology, Warsaw 00-661, Poland}
\affiliation{Wayne State University, Detroit, Michigan 48201}
\affiliation{Wuhan University of Science and Technology, Wuhan, Hubei 430065}
\affiliation{Yale University, New Haven, Connecticut 06520}

\author{B.~E.~Aboona}\affiliation{Texas A\&M University, College Station, Texas 77843}
\author{J.~Adam}\affiliation{Czech Technical University in Prague, FNSPE, Prague 115 19, Czech Republic}
\author{G.~Agakishiev}\affiliation{Joint Institute for Nuclear Research, Dubna 141 980}
\author{I.~Aggarwal}\affiliation{Panjab University, Chandigarh 160014, India}
\author{M.~M.~Aggarwal}\affiliation{Panjab University, Chandigarh 160014, India}
\author{Z.~Ahammed}\affiliation{Variable Energy Cyclotron Centre, Kolkata 700064, India}
\author{A.~Aitbayev}\affiliation{Joint Institute for Nuclear Research, Dubna 141 980}
\author{I.~Alekseev}\affiliation{Alikhanov Institute for Theoretical and Experimental Physics NRC "Kurchatov Institute", Moscow 117218}\affiliation{National Research Nuclear University MEPhI, Moscow 115409}
\author{E.~Alpatov}\affiliation{National Research Nuclear University MEPhI, Moscow 115409}
\author{A.~K.~Alshammri}\affiliation{Kent State University, Kent, Ohio 44242}
% ...

\collaboration{STAR Collaboration}\noaffiliation

\maketitle

\section{3-Gaussian Fit}

\begin{figure}[htb!]
    \vspace{-5pt}
    \centering
    \includegraphics[width=6cm]{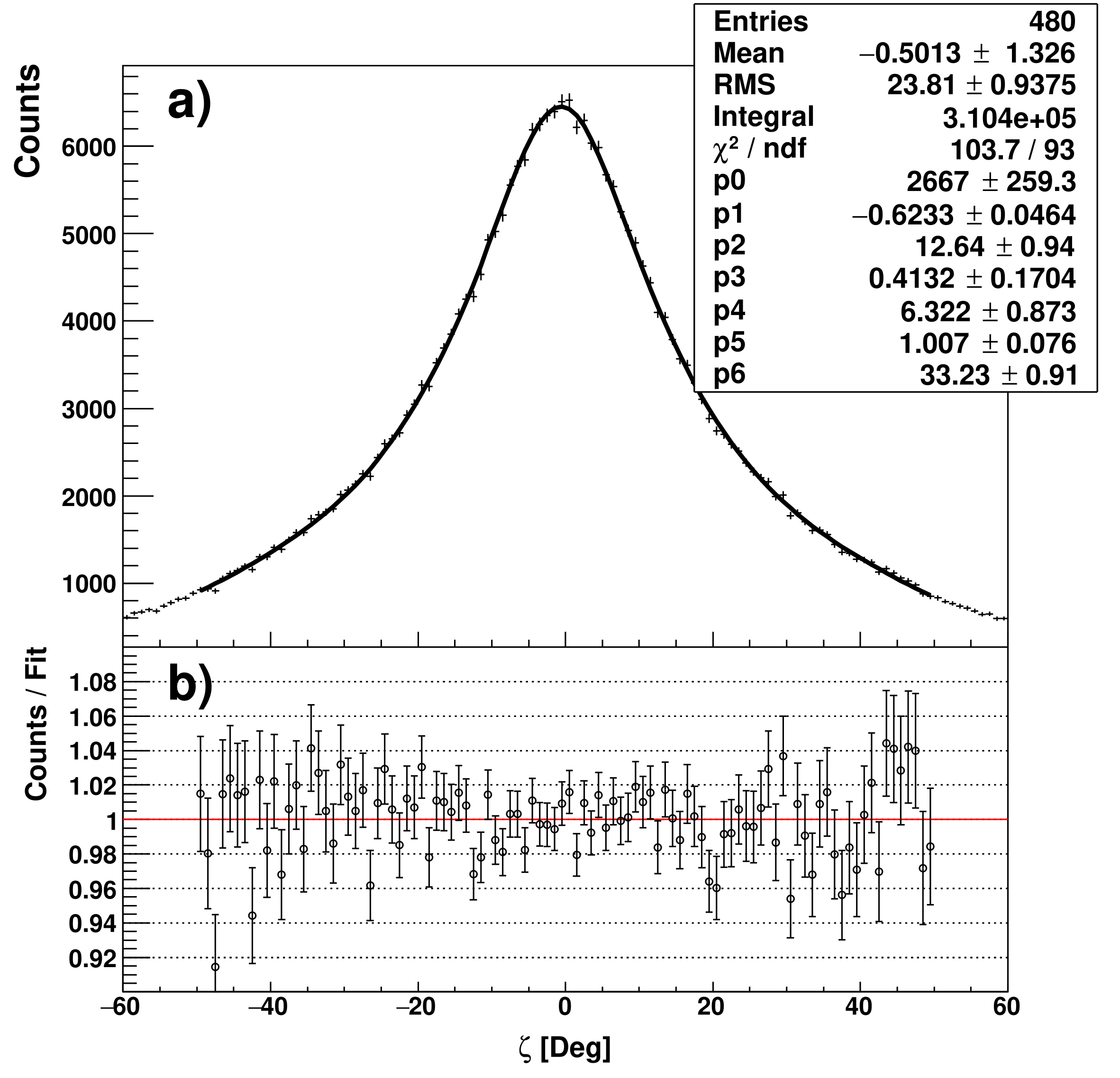}
    \vspace{-10pt}
    \caption{Distribution of $\zeta$ in the range of $\phi_b$ $\in$ [15$^\circ$, 30$^\circ$] and $\etatot$ $\in$ [0, 0.8] of the unpolarized +tagging data for the +$z$ beam. A 3-Gaussian fit is performed within the range $\pm$50$^\circ$. The lower panel shows a ratio of counts over the fit.
    }
    \vspace{-10pt}
    \label{supplement_fig1}
\end{figure}

An example of a 3-Gaussian fit (
    $N(\zeta)$ =
    $p_0 \cdot (e^{-\frac{{(\zeta - p_1)}^2}{2 {p_2}^2}}$ 
    + $p_3 \cdot e^{-\frac{{(\zeta - p_1)}^2}{2 {p_4}^2}}$
    + $p_5 \cdot e^{-\frac{{(\zeta - p_1)}^2}{2 {p_6}^2}})$
) of the N($\zeta$) distribution
for $\zeta$ within $\phi_b$ $\in$ [15$^\circ$, 30$^\circ$] and $\etatot$ $\in$ [0, 0.8] using unpolarized +tagged +$z$ beam data is shown in Fig.~\ref{supplement_fig1}.

The triple Gaussian shape 
is chosen for the $\zeta$ fitting 
because it captures the salient features of the $\zeta$ distribution. The distribution itself 
is driven by broad (Gaussian) widths arising largely from multi-GeV intrinsic 
("primordial") ${k_T}$ and 
initial-state radiation (ISR) effects, in 
convolution with various responses
arising from the STAR detector readout and triggering. 
These features may be further shaped by the influence of
jet-patch triggering thresholds, supplying events with
different average $p_T$. However, there is no detailed
accounting of contributing components, nor the relevance or role played by the several
possible additional modifications. In this context, it is most important to again note, 
that our extracted dijet Sivers $\dzeta$ effect results are largely insensitive 
to variations of the empirically-driven Gaussian function shape choice.

\section{$\dzeta$ Results}

The primary experimental result of this analysis for the Sivers effect, is the 
dijet spin-dependent "tilt" we measure as a shift $\dzeta$ of the opening angle centroid. A compilation of these numerical values is given in  
Table~\ref{tab:dZeta_results}, where the single-spin results from the two colliding beams are combined as described in the main text.

\begin{table}[htb!]
    \centering
    \resizebox{\columnwidth}{!}{
    \begin{tabular}{|c | c | c | c | c |}
      \hline 
	  $\langle \etatot \rangle$ & \textbf{+} tag (deg) & \textbf{-} tag (deg) & \textbf{$0^{+}$} tag (deg) & \textbf{$0^{-}$} tag (deg)\\
	  \hline 
	  $[$-3.6, -2.6$]$ & -0.302$\substack{\pm0.253 \\ \pm0.017}$ & -0.072$\substack{\pm0.228 \\ \pm0.004}$ & -0.384$\substack{\pm0.223 \\ \pm0.022}$ & 0.030$\substack{\pm0.256 \\ \pm0.002}$ \\
	  $[$-2.6, -1.6$]$ & 0.015$\substack{\pm0.071 \\ \pm0.001}$ & 0.028$\substack{\pm0.080 \\ \pm0.002}$ & 0.035$\substack{\pm0.063 \\ \pm0.002}$ & 0.050$\substack{\pm0.070 \\ \pm0.003}$ \\
	  $[$-1.6, -0.8$]$ & -0.003$\substack{\pm0.044 \\ \pm0.004}$ & -0.072$\substack{\pm0.050 \\ \pm0.005}$ & 0.016$\substack{\pm0.038 \\ \pm0.003}$  & -0.014$\substack{\pm0.043 \\ \pm0.003}$ \\
	  $[$-0.8, 0$]$ & 0.034$\substack{\pm0.034 \\ \pm0.004}$ & -0.055$\substack{\pm0.038 \\ \pm0.004}$& -0.034$\substack{\pm0.029 \\ \pm0.004}$ & -0.058$\substack{\pm0.033 \\ \pm0.004}$ \\
	  $[$0, 0.8$]$ & 0.009$\substack{\pm0.033 \\ \pm0.003}$ & -0.058$\substack{\pm0.038 \\ \pm0.004}$ & 0.029$\substack{\pm0.029 \\ \pm0.003}$& 0.022$\substack{\pm0.033 \\ \pm0.002}$ \\
	  $[$0.8, 1.6$]$ & 0.088$\substack{\pm0.042 \\ \pm0.006}$& -0.060$\substack{\pm0.051 \\ \pm0.004}$ & 0.041$\substack{\pm0.036 \\ \pm0.004}$ & -0.025$\substack{\pm0.044 \\ \pm0.003}$ \\
	  $[$1.6, 2.6$]$ & 0.075$\substack{\pm0.072 \\ 0.006}$ & -0.033$\substack{\pm0.087 \\ \pm0.003}$ & -0.010$\substack{\pm0.054 \\ \pm0.003}$ & -0.048$\substack{\pm0.073 \\ \pm0.0003}$ \\
	  $[$2.6, 3.6$]$ & 0.096$\substack{\pm0.282 \\ \pm0.008}$ & -0.583$\substack{\pm0.342 \\ \pm0.033}$ & 0.084$\substack{\pm0.178 \\ \pm0.006}$ & -0.067$\substack{\pm0.264 \\ \pm0.005}$ \\
	  \hline
    \end{tabular}
    }
    \caption{The measured average centroid shift, $\dzeta$ in degrees, for the jet-charge selections
    plus, minus, $0^+$ and $0^-$ vs. $\langle \etatot \rangle$ for the combined $+z$- and $-z$-going beams.
    For each entry, the upper indicated error is statistical, while the lower one is
    systematic (largely fit range uncertainty and smearing effects).}
    \label{tab:dZeta_results}
\end{table}

\section{$\dzeta$ vs. $A_N$}

Our analysis chose the observable $\dzeta$ as the means of characterizing the Sivers dijet asymmetry in the present study. A summation of counts to the "left" vs. "right" in the $\zeta$ opening angle distribution, thus constructing an $A_N$, was the choice in our initial
published Sivers dijet measurement~\cite{2006dijet} and has been used in recent theoretical calculations~\cite{PhysRevD.102.114012,2021JHEP...02..066K}. The two approaches are related, as they refer to the same basic dijet Sivers asymmetry. Here, since we use Gaussians to characterize the $\zeta$ distribution,
the asymmetries can be related analytically: e.g., comparison of the area of a single Gaussian to the nearly rectangular area
swept out by a small shift in $\zeta$, gives the relation 
$\Delta\langle\zeta\rangle = A_N \times (\sigma\times \sqrt{2\pi})$, where $\sigma$ is
the width of the Gaussian, with a straight-forward extension to multiple Gaussians
(here, the ratio of the sums of the single contributions from the three Gaussian components). From a study of the fitting parameters of the saved Gauss fits 
representing the totality of the analyzed data, we deduce a
conservative estimate of the numerical relationship $A_N$= Factor*$\dzeta$, where the Factor = 0.0223 with a 0.0023 uncertainty. The uncertainty is dominated by the RMS error of the Factor calculation from the numerous underlying 3-Gauss fits, but also includes a small contribution from a slight $\etatot$ dependence. This of course doesn't replace a full $A_N$ type analysis, but does provide a meaningful bridge for comparison to the theoretical approaches, while future experimental analyses may more fully include $A_N$ results.

 \section{Cosine Fit}

\begin{figure}[htb!]
    \vspace{-5pt}
    \centering
    \includegraphics[width=7cm]{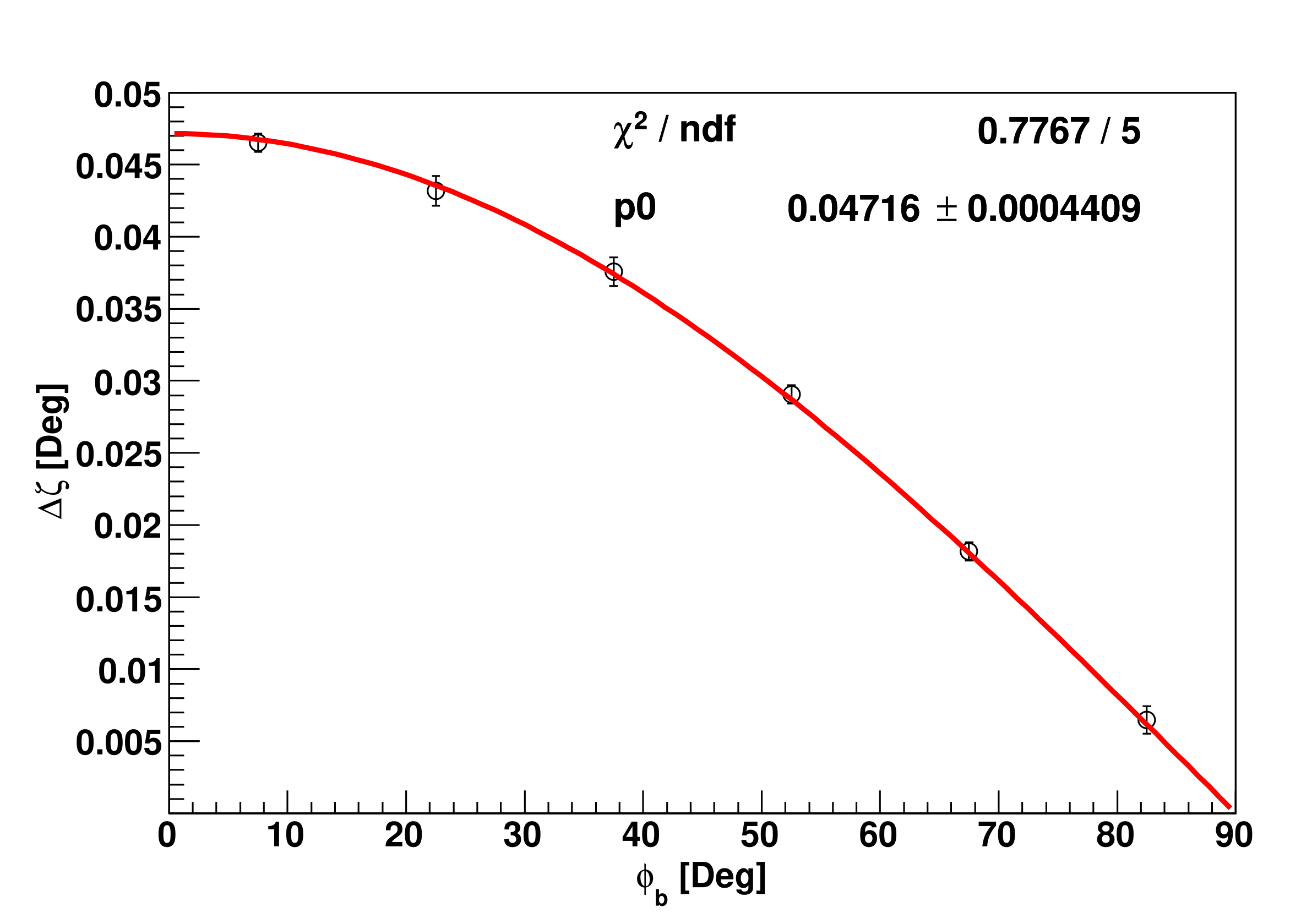}
    \vspace{-10pt}
    \caption{Example of a Cosine fit with $\kt$ = 5 MeV/c manually inserted in corrected data for $\etatot$ $\in$ [0, 0.8].}
    \vspace{-10pt}
    \label{supplement_fig2}
\end{figure}

An example of a cosine fit ($\dzeta$($\phi_b$) = $p_0$ $\cdot$ $cos(\phi_b)$) 
for $\dzeta$-vs-$\phi_b$ in the range of $\etatot$ $\in$ [0, 0.8] is shown in Fig.~\ref{supplement_fig2}. In the toy model, a constant 5 MeV/c $\kt$ is separately inserted in the corrected data along the +$x$ and $-x$ directions to mimic a Sivers effect in the spin-up and spin-down states, respectively, in order to "calibrate" the $\dzeta$ vs.
$\kt$ kinematic relationship.

\section{Parton Fractions}

The overall Sivers dijet effect comes from four contributions: 
$u$, $d$, gluons, and sea quarks. In our analysis, since the sea quark 
fraction is relatively low and their overall contribution is expected 
to be small, its contribution is combined with that of the gluons
Along with the four charge tagged $\kt$ results, there are then enough 
constraints to solve for the $\kt$ of the three parton sources.

A requirement for obtaining the solutions is the parton fractions 
in each charge tagged bin, which can be estimated from the embedding.
Table~\ref{tab:purity_comb} shows the parton fractions after combining 
all the data sets (JP and run), as well as combining results from both the $+ z$ going (Blue) and $- z$ going 
(Yellow) beams. The $u$ quark fraction sequentially decreases from the 
plus-tagging to $0^+$, $0^-$, and the minus-tagging, while the $d$ quark 
fraction increases. The quark fraction, particularly the $u$ quark fraction, 
is positively correlated with $\etasum$, which is proportional to 
$log(x_{1}/x_{2})$. To the contrary, the gluon fraction has a negative 
correlation with $\etasum$. Because of the statistical uncertainties associated with 
the inversion, these effects are not strongly manifest in our results for
partonic $\kt$.

\begin{table}[phtb!]
    \renewcommand{\arraystretch}{1.4}
    \centering
    \begin{tabular}{|c|c|c|c|c|c|}
      \hline 
      \multirow{8}{*}{$\substack{\textbf{+} \\ tag}$} & $\langle\etasum\rangle$ &  [-3.6, -2.6] & [-2.6, -1.6] & [-1.6, -0.8] & [-0.8, 0]  \\
      \cline{2-6} 
      & $f^{u}$ & 0.047$\substack{+0.046 \\ -0.029}$ & 0.154$\substack{+0.032 \\ -0.031}$ & 0.239$\substack{+0.040 \\ -0.040}$ & 0.313$\substack{+0.045 \\ -0.045}$ \\
      \cline{2-6}
      & $f^{d}$ & 0.175$\substack{+0.072 \\ -0.061}$ & 0.061$\substack{+0.015 \\ -0.014}$ & 0.077$\substack{+0.013 \\ -0.013}$ & 0.086$\substack{+0.013 \\ -0.013}$ \\
      \cline{2-6} 
      & $f^{g+sea}$ & 0.778$\substack{+0.139 \\ -0.100}$ & 0.785$\substack{+0.045 \\ -0.044}$ & 0.684$\substack{+0.041 \\ -0.041}$ & 0.601$\substack{+0.045 \\ -0.045}$ \\
      \cline{2-6}
      & $\langle\etasum\rangle$  & [0, 0.8] & [0.8, 1.6] & [1.6, 2.6] & [2.6, 3.6]      \\
      \cline{2-6}
      & $f^{u}$ & 0.433$\substack{+0.044 \\ -0.044}$ & 0.578$\substack{+0.049 \\ -0.049}$ & 0.670$\substack{+0.048 \\ -0.048}$ & 0.798$\substack{+0.082 \\ -0.097}$ \\
      \cline{2-6}
      & $f^{d}$ & 0.109$\substack{+0.014 \\ -0.014}$ & 0.112$\substack{+0.015 \\ -0.015}$ & 0.128$\substack{+0.021 \\ -0.020}$ & 0.175$\substack{+0.088 \\ -0.071}$ \\
      \cline{2-6}
      & $f^{g+sea}$ & 0.458$\substack{+0.038 \\ -0.038}$ & 0.311$\substack{+0.036 \\ -0.035}$ & 0.202$\substack{+0.039 \\ -0.038}$ & 0.027$\substack{+0.152 \\ -0.019}$ \\
      \hline 
      \hline 
      \multirow{8}{*}{$\substack{\textbf{-} \\ tag}$} & $\langle\etasum\rangle$ &  [-3.6, -2.6] & [-2.6, -1.6] & [-1.6, -0.8] & [-0.8, 0]  \\
      \cline{2-6}
      & $f^{u}$ &  0.048$\substack{+0.039 \\ -0.026}$ & 0.060$\substack{+0.015 \\ -0.014}$ & 0.089$\substack{+0.016 \\ -0.016}$ & 0.141$\substack{+0.024 \\ -0.024}$  \\
      \cline{2-6}
      & $f^{d}$ & 0.061$\substack{+0.042 \\ -0.030}$ & 0.107$\substack{+0.025 \\ -0.024}$ & 0.161$\substack{+0.026 \\ -0.026}$ & 0.178$\substack{+0.025 \\ -0.025}$ \\
      \cline{2-6}
      & $f^{g+sea}$ & 0.891$\substack{+0.100 \\ -0.075}$ & 0.833$\substack{+0.040 \\ -0.038}$ & 0.750$\substack{+0.034 \\ -0.034}$ & 0.681$\substack{+0.037 \\ -0.036}$ \\
      \cline{2-6}
      & $\langle\etasum\rangle$  & [0, 0.8] & [0.8, 1.6] & [1.6, 2.6] & [2.6, 3.6]      \\
      \cline{2-6}
      & $f^{u}$ & 0.200$\substack{+0.030 \\ -0.030}$ & 0.272$\substack{+0.038 \\ -0.038}$ & 0.406$\substack{+0.062 \\ -0.061}$ & 0.466$\substack{+0.137 \\ -0.134}$ \\
      \cline{2-6}
      & $f^{d}$ & 0.258$\substack{+0.034 \\ -0.034}$ & 0.268$\substack{+0.035 \\ -0.035}$ & 0.350$\substack{+0.065 \\ -0.064}$ & 0.170$\substack{+0.117 \\ -0.086}$ \\
      \cline{2-6}
      & $f^{g+sea}$ & 0.542$\substack{+0.039 \\ -0.039}$ & 0.461$\substack{+0.048 \\ -0.047}$ & 0.244$\substack{+0.069 \\ -0.055}$ & 0.364$\substack{+0.290 \\ -0.145}$ \\
      \hline       
      \hline 
      \multirow{8}{*}{$\substack{\textbf{$0^{+}$} \\ tag}$} & $\langle\etasum\rangle$ &  [-3.6, -2.6] & [-2.6, -1.6] & [-1.6, -0.8] & [-0.8, 0]  \\
      \cline{2-6}
      & $f^{u}$ &  0.151$\substack{+0.063 \\ -0.055}$ & 0.110$\substack{+0.028 \\ -0.027}$ & 0.191$\substack{+0.035 \\ -0.035}$ & 0.239$\substack{+0.033 \\ -0.033}$   \\
      \cline{2-6}
      & $f^{d}$ & 0.031$\substack{+0.034 \\ -0.021}$ & 0.079$\substack{+0.021 \\ -0.020}$ & 0.101$\substack{+0.019 \\ -0.019}$ & 0.132$\substack{+0.019 \\ -0.019}$ \\
      \cline{2-6}
      & $f^{g+sea}$ & 0.818$\substack{+0.106 \\ -0.080}$ & 0.811$\substack{+0.047 \\ -0.044}$ & 0.708$\substack{+0.039 \\ -0.039}$ & 0.629$\substack{+0.036 \\ -0.036}$ \\
      \cline{2-6}
      & $\langle\etasum\rangle$  & [0, 0.8] & [0.8, 1.6] & [1.6, 2.6] & [2.6, 3.6]      \\
      \cline{2-6}
      & $f^{u}$ & 0.360$\substack{+0.043 \\ -0.043}$ & 0.448$\substack{+0.046 \\ -0.046}$ & 0.612$\substack{+0.049 \\ -0.049}$ & 0.663$\substack{+0.062 \\ -0.063}$ \\
      \cline{2-6}
      & $f^{d}$ & 0.152$\substack{+0.021 \\ -0.020}$ & 0.177$\substack{+0.027 \\ -0.027}$ & 0.179$\substack{+0.026 \\ -0.026}$ & 0.194$\substack{+0.046 \\ -0.044}$ \\
      \cline{2-6}
      & $f^{g+sea}$ & 0.487$\substack{+0.038 \\ -0.038}$ & 0.375$\substack{+0.042 \\ -0.042}$ & 0.209$\substack{+0.037 \\ -0.036}$ & 0.144$\substack{+0.052 \\ -0.042}$ \\
      \hline
      \hline 
      \multirow{8}{*}{$\substack{\textbf{$0^{-}$} \\ tag}$} & $\langle\etasum\rangle$ &  [-3.6, -2.6] & [-2.6, -1.6] & [-1.6, -0.8] & [-0.8, 0]  \\
      \cline{2-6}
      & $f^{u}$ & 0.036$\substack{+0.029 \\ -0.020}$ & 0.082$\substack{+0.022 \\ -0.020}$ & 0.117$\substack{+0.023 \\ -0.022}$ & 0.201$\substack{+0.028 \\ -0.028}$  \\
      \cline{2-6}
      & $f^{d}$ & 0.109$\substack{+0.049 \\ -0.042}$ & 0.079$\substack{+0.020 \\ -0.018}$ & 0.133$\substack{+0.025 \\ -0.024}$ & 0.153$\substack{+0.022 \\ -0.022}$ \\
      \cline{2-6}
      & $f^{g+sea}$ & 0.855$\substack{+0.087 \\ -0.070}$ & 0.838$\substack{+0.045 \\ -0.043}$ & 0.750$\substack{+0.041 \\ -0.041}$ & 0.646$\substack{+0.037 \\ -0.037}$ \\
      \cline{2-6}
      & $\langle\etasum\rangle$  & [0, 0.8] & [0.8, 1.6] & [1.6, 2.6] & [2.6, 3.6]      \\
      \cline{2-6}
      & $f^{u}$ & 0.279$\substack{+0.034 \\ -0.034}$ & 0.373$\substack{+0.039 \\ -0.039}$ & 0.530$\substack{+0.046 \\ -0.046}$ & 0.659$\substack{+0.078 \\ -0.081}$ \\
      \cline{2-6}
      & $f^{d}$ & 0.198$\substack{+0.026 \\ -0.026}$ & 0.217$\substack{+0.033 \\ -0.033}$ & 0.211$\substack{+0.033 \\ -0.033}$ & 0.214$\substack{+0.067 \\ -0.061}$ \\
      \cline{2-6}
      & $f^{g+sea}$ & 0.523$\substack{+0.040 \\ -0.040}$ & 0.410$\substack{+0.046 \\ -0.046}$ & 0.259$\substack{+0.047 \\ -0.046}$ & 0.127$\substack{+0.084 \\ -0.047}$ \\
      \hline 
    \end{tabular}
    \caption{Parton fraction and statistical uncertainty after Blue and Yellow beam combination for individual partons in the four tagged bins. The JP2 and JP1 triggers in the 2012 embedding and the 2015 embedding are all combined together.}
    \label{tab:purity_comb}
\end{table}

The systematic errors associated with the parton fraction calculations arise from several sources in the simulation.
The largest contributing factors to the uncertainty are the PDF choice and initial/final 
state radiation (ISR/FSR), as well as the statistics of the simulation sample. 
Different PDF sets directly cause discrepancies in the fraction of partons.
The amount of ISR/FSR particularly affects event selection in the low $p_T$ region, 
which leads to uncertainties in the parton fractions.
These uncertainties due to PDF and ISR/FSR are estimated by varying respective PYTHIA tunes, 
comparing to the default tune (370) and quoting the average absolute difference.
The statistical uncertainties of parton fractions are at about the same level as 
the PDF and ISR/FSR uncertainties, and are added in quadrature to the total systematics. 
In total these uncertainties vary with parton purity in the 
various 
charge bins and as a function of 
$-3.6 < \etatot < 3.6$, ranging from 18 to 7-12\% for $u$ and $d$, and 3-21\% for the $g+sea$.  
The combined effect of these and other systematic errors on the extracted parton $\kt$ values are collected and summarized in Table~\ref{tab:syst_err} below.

\section{Individual parton $\kt$ Results}

The numerical values, including the statistical and systematic errors, for the bin-by-bin parton $\kt$ values are listed in 
Table~\ref{tab:finalresult_kt_inv}.

\begin{table}[phtb!]
   \resizebox{\columnwidth}{!}{
    \begin{tabular}{c c c c c}
     \hline 
     \hline
     \multicolumn{5}{c}{\textbf{The average parton $\langle x \rangle$ and $\kt$ [MeV/c]}} \\ 
    $\etatot$ & [-3.6, -1.6] & [-1.6, 0] & [0, 1.6] & [1.6, 3.6] \\
      \hline
      $\langle x^u \rangle$ & 0.05$\pm$0.02 & 0.11$\pm$0.05 & 0.22$\pm$0.09 & 0.36$\pm$0.11 \\
      $\ktu$ [MeV/c] & -37.5$\pm$132.3 & 11.0$\pm$20.2 & 19.8$\pm$9.7 & 28.0$\pm$21.9  \\
      \hline    
      $\langle x^d \rangle$ & 0.04$\pm$0.02 & 0.11$\pm$0.05 & 0.20$\pm$0.08 & 0.33$\pm$0.09 \\
      $\ktd$ [MeV/c] & -50.5$\pm$236.3 & -91.6$\pm$53.3 & -34.3$\pm$31.8 & 12.3$\pm$63.9 \\
      \hline    
      $\langle x^{g+sea} \rangle$ & 0.04$\pm$0.02 & 0.09$\pm$0.04 & 0.17$\pm$0.07 & 0.30$\pm$0.09 \\
      $\ktgsea$ [MeV/c] & 11.5$\pm$36.6 & 10.5$\pm$13.7 & -0.0$\pm$16.1 & -88.7$\pm$89.8 \\
      \hline 
      \hline
    \end{tabular}
    }
    \caption{The average $x$ and inverted parton $\kt$ for $u$, $d$, and $g$+sea in each $\etatot$ region. The rms of $x$ and the total uncertainty of $\kt$ are also given in the table.}
    \vspace{-10pt}
    \label{tab:finalresult_kt_inv}
\end{table}

\section{Total Systematic Uncertainties}

The total systematic uncertainty and each individual contribution for the bin-by-bin and average parton $\kt$ values are listed in Table~\ref{tab:syst_err}. 

\newpage

\begin{table}[phtb!]
   \resizebox{\columnwidth}{!}{
    \begin{tabular}{c l c c c c c}
     \hline
     \hline
     \multicolumn{7}{c}{\textbf{Systematic uncertainties of parton $\kt$ [MeV/c]}} \\
    \multicolumn{2}{c}{$\etatot$} & [-3.6, -1.6] & [-1.6, 0] & [0, 1.6] & [1.6, 3.6] & Avg \\
     \hline
    \multirow{4}{*}{$u$} & Simulation  & 49.6 &	6.4	&	2.6	&	8.3	&	2.3 \\
    & Fitting     &   15.3	&	2.4	&	1.1	&	5.9	&	1.0 \\
    & Measurement &   3.7   &	1.1	&	0.8	&	0.8	&	0.5 \\
     \cline{2-7}
    & Total     &   52.0	&	6.9	&	3.0	&	10.2	&	2.6 \\
     \hline
    \multirow{4}{*}{$d$} & Simulation  &   79.7	&	23.4	&	9.8	&	29.2	&		8.6 \\
    & Fitting     &   28.3	&	6.1	&	3.8	&	15.5	&		3.1 \\
    & Measurement &   12.6	&	2.9	&	1.5	&	2.2	&		1.1 \\
     \cline{2-7}
    & Total      &   85.5	&	24.3    &	10.7	&	33.1	&		9.3 \\
     \hline   
    \multirow{4}{*}{$g$+sea} & Simulation  &   13.1	&	5.5	&	4.9	&	41.4	&		3.5 \\
    & Fitting     & 4.4	&	1.6	&	1.9	&	22.0	&		1.2 \\
    & Measurement & 1.7	&	0.6	&	0.9	&	3.5	&		0.5 \\
     \cline{2-7} 
    & Total      & 13.9	&	5.8	&	5.4	&	47.0	&		3.8 \\    
     \hline
     \hline
    \end{tabular}
    }    
    \caption{The total systematic uncertainty [MeV/c] of each parton $\kt$ and breakdown of individual contributions for each $\etatot$ region and the average of all $\etatot$ regions.}
    \vspace{-10pt}
    \label{tab:syst_err}
\end{table}

\section*{Acknowledgements}
We thank the RHIC Operations Group and SCDF at BNL, the NERSC Center at LBNL, and the Open Science Grid consortium for providing resources and support.  This work was supported in part by the Office of Nuclear Physics within the U.S. DOE Office of Science, the U.S. National Science Foundation, National Natural Science Foundation of China, Chinese Academy of Science, the Ministry of Science and Technology of China and the Chinese Ministry of Education, NSTC Taipei, the National Research Foundation of Korea, Czech Science Foundation and Ministry of Education, Youth and Sports of the Czech Republic, Hungarian National Research, Development and Innovation Office, New National Excellency Programme of the Hungarian Ministry of Human Capacities, Department of Atomic Energy and Department of Science and Technology of the Government of India, the National Science Centre and WUT ID-UB of Poland, German Bundesministerium f\"ur Bildung, Wissenschaft, Forschung and Technologie (BMBF), Helmholtz Association, Ministry of Education, Culture, Sports, Science, and Technology (MEXT), Japan Society for the Promotion of Science (JSPS), and Agencia Nacional de Investigacion y Desarrollo de Chile (ANID), Chile.

\bibliographystyle{apsrev4-1} 
\bibliography{suppl_text}